\documentclass[amsmath,amssymb,floatfix,preprint,prl]{revtex4-1}
\usepackage{graphicx}
\usepackage{bm}
\usepackage[]{units}
\usepackage{hyperref}
\usepackage{dcolumn}
\usepackage[caption=false]{subfig}

\begin{document}

\newcommand{\etal}{\MakeLowercase{\textit{et al.}}} 

\title{A generalized self-veto probability for atmospheric neutrinos}

\preprint{Preprint submitted to Physical Review Letters}

\author{Thomas K. Gaisser}
\address{Bartol Research Institute and Dept.~of Physics and Astronomy, University of Delaware, Newark, DE 19716, USA}
\author{Kyle Jero}
\affiliation{Dept.~of Physics and Wisconsin IceCube Particle Astrophysics Center, University of Wisconsin, Madison, WI 53706, USA}
\author{Albrecht Karle}
\address{Dept.~of Physics and Wisconsin IceCube Particle Astrophysics Center, University of Wisconsin, Madison, WI 53706, USA}
\author{Jakob van Santen}
\email{jvansanten@icecube.wisc.edu}
\address{Dept.~of Physics and Wisconsin IceCube Particle Astrophysics Center, University of Wisconsin, Madison, WI 53706, USA}

\begin{abstract}
Neutrino telescopes such as IceCube 
search for an excess of high energy neutrinos above
the steeply falling atmospheric background as one
approach to finding extraterrestrial neutrinos.  For 
samples of events selected to start in the detector,
the atmospheric background can be reduced to the extent
that a neutrino interaction inside the fiducial volume
is accompanied by a detectable muon from the same cosmic-ray
cascade in which the neutrino was produced.  Here we provide
an approximate calculation of the veto probability as a function
of neutrino energy and zenith angle.
\end{abstract}

\maketitle

\section{Introduction}
\label{sec1}

A downward atmospheric neutrino will be excluded from a sample
of atmospheric neutrinos if the sample consists of events starting
in the detector and if the neutrino has sufficiently high energy
and sufficiently small zenith angle that a muon from the same event
will enter the detector at the same time and be recognized.  Such an event will
be classified as an atmospheric muon and rejected.  This strategy was
originally suggested in \cite{Schoenert:2009} for the case of $\nu_\mu$ from
the decays of charged pions and kaons, where the probability that the muon
produced in the same decay reaches the detector can be calculated with an
analytic approximation. A preliminary estimate of the more general case where
the veto is provided by any muon from the same shower was used in evaluating
the atmospheric neutrino background in the high energy starting event (HESE)
analysis of IceCube~\cite{HESE}.

In this paper we show how to calculate the more general case in which the
neutrino can be accompanied by a muon produced in any branch of the same
shower. Accounting for these extra muons increases the veto probability for
$\nu_\mu$ only slightly. For $\nu_e$ however, such uncorrelated muons are the
only source of accompanying muons since meson decays to electron neutrinos are
accompanied by electrons rather than muons. Treating $\nu_e$ properly is
especially important when considering neutrino fluxes from charmed mesons,
which decay to $\nu_e$ and $\nu_\mu$ with nearly equal probability.

Calculating the probability that a neutrino is accompanied by a muon produced
in any branch of the same shower requires a calculation that accounts for the
correlation between the candidate neutrino and the entire shower structure.
This is a straightforward Monte Carlo calculation that is limited only by the
statistics of meson decay at high energy. The characteristic ratio of decay
probability to interaction probability for a meson of type $\alpha$ is
$\epsilon_\alpha/(E_\nu\cos\theta)$. At energies of $100$~TeV and above, of
current interest in IceCube, this ratio for kaons is significantly less than
1\%. For prompt neutrinos from charm decay, the ratio is by definition large,
but the production of charmed hadrons itself is rare and subject to large
uncertainties. For these reasons it is useful to develop a numerical estimate
of the probability that a neutrino is accompanied by an unrelated muon from the
same event. If such an approximation can be shown to agree with the Monte Carlo
result at low energy, it can be used to extend the veto calculation beyond the
statistical limitations of the full Monte Carlo.

\section{Calculations}

The flux of atmospheric neutrinos can be
obtained by integrating the production spectrum of neutrinos over atmospheric depth.
The production spectrum is an integral over the parent spectrum of mesons
that decay to produce neutrinos.  The range of the integration is given
by the maximum and minimum kinematically allowed values of $E_{\rm parent}$ for
a given $E_\nu$.  
For a power-law primary spectrum of nucleons, the integral over the 
neutrino production spectrum leads to the standard approximation~\cite{Gaisser:1990vg}
for the flux of $\nu_\mu+\bar{\nu}_\mu$ from decay of charged pions and kaons:
\begin{align}
\nonumber & \phi_\nu(E_\nu)\, = \,\phi_N(E_\nu) \\ & \times
 \left\{{A_{\pi\nu}\over 1 +
 B_{\pi\nu}\cos\theta\, E_\nu / \epsilon_\pi}
 + {A_{K\nu}\over 1+B_{K\nu}\cos\theta\, E_\nu / \epsilon_K}\right\}.
\label{angular}
\end{align}

In the case of the two body decays of charged pions and charged kaons, the
integral over the parent meson energy can be constrained to require 
\begin{equation}
E_\mu\,+\,E_\nu\,>\,E_{\mu,{\rm min}}\,+\,E_\nu,
\label{Emumin}
\end{equation}
where $E_{\mu,{\rm min}}$ is the minimum muon energy needed to
reach the depth of the detector and trigger it.
This is the calculation of Ref.~\cite{Schoenert:2009} which leads to
a modified $\phi_\nu^*(E_\nu,E_{\mu,{\rm min}})$, which is
the flux of neutrinos accompanied by the muon from the same meson
decay.  Then the passing rate,
\begin{equation}
P(E_\nu,E_{\mu,{\rm min}})\,=\,\frac{\phi_\nu(E_\nu)\,-\,\phi_\nu^*(E_\nu,E_{\mu,{\rm min}})}{\phi_\nu(E_\nu)},
\label{passing}
\end{equation}
gives the fraction of atmospheric $\nu_\mu$ that are not accompanied by the muon from
the same decay in which the neutrino was produced.
Our goal is to generalize the passing rate to atmospheric neutrinos of
all flavors by including all muons produced in the same 
cosmic ray shower as the neutrino.

\begin{figure}[h]
	\centering
	\subfloat[Cumulative lepton yields for vertical, 1 PeV proton showers.]{
	    \includegraphics[scale=1]{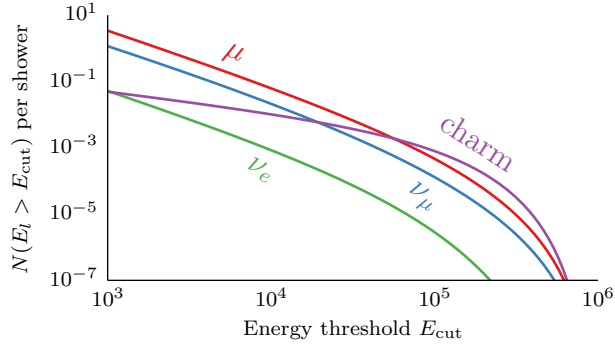}
		\label{fig:elbert:families}
	}
	
	\subfloat[Zenith angle dependence of the cumulative conventional $\nu_{e}$
	yield evaluated at 10\% of the primary energy for different primary energies.]{
	    \includegraphics[scale=1]{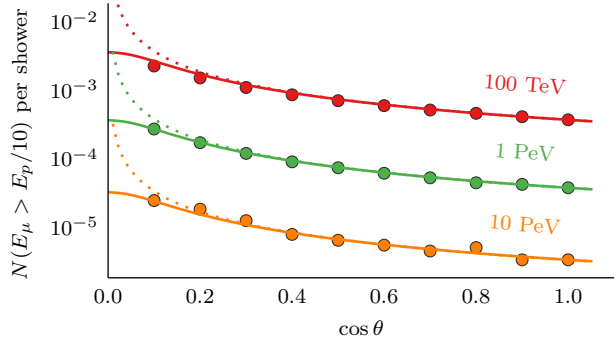}
		\label{fig:elbert:zenith}
	}
	\caption{(Color online)
	Lepton yields from the modified Elbert formula.
	Each curve in \protect\subref{fig:elbert:families} shows Eq.~\eqref{Elbert}
	evaluated with one of the parameter sets from Table~\ref{tab:elbert_params}.
	The points in \protect\subref{fig:elbert:zenith} show yields from \textsc{corsika}~\cite{CORSIKA}
	simulation, the dotted lines a $1/\cos\theta$ dependence, and the solid
	lines a $1/\cos^*\theta$ dependence \cite{Chirkin:2004ic}. All conventional
	lepton yields have the same zenith dependence.
	}
	\label{fig:elbert}
\end{figure}

One approach to a numerical evaluation of the passing rate
is to use an approximate form for the yield of muons per primary nucleus.
We use an approximation based on parameterization of simulations that is
sometimes referred to as the Elbert formula~\cite{Elbert}.  This
approximation gives a good description of the average properties of
muon bundles generated by primary cosmic ray nuclei of mass $A$ and total energy $E$ ~\cite{Gaisser:1985yw}.
(See also Ref.~\cite{Lipari:1993ty}.)
We find, using the simulations described in the results section,
that the Elbert formula can be generalized to describe fluxes of $\nu_{\mu}$ and $\nu_e$
by adding one additional parameter.
In integral form, the approximation is
\begin{equation}
N_l(>E_l,A,E,\theta) = {{\rm K}_l}\,\frac{A}{E_l\cos^*\theta}x^{-p_1}
\left(1-x^{p_3}\right)^{p_2},
\label{Elbert}
\end{equation}
with $x \equiv A E_l / E$ and the constants K, $p_1$, $p_2$, and $p_3$
given for different leptons $l$ in Table~\ref{tab:elbert_params}.
The approximation is valid above a few TeV, where pions and kaons are
more likely to interact than decay in flight. The decay probability is
proportional to $1 / E_l \cos^*\theta$~\footnote{$\cos^*\theta$ is $\cos\theta$ 
evaluated at the average altitude of the first interaction parameterized as
in Ref.~\cite{Chirkin:2004ic}. The correction is important for zenith angles
larger than 70$^{\circ}$.}. The same form can be made to describe leptons from
the decays of charmed mesons like the $D^{\pm}$ that decay promptly before
they can re-interact by removing the decay-probability factor:
\begin{equation}
N_l(>E_l,A,E,\theta)\,=\,{{\rm K}_l}\, A x^{-p_1}
\left(1-x^{p_3}\right)^{p_2}.
\label{ElbertCharm}
\end{equation}
Fig.~\ref{fig:elbert} shows the approximate lepton yields as a function of
lepton energy, primary energy, and zenith angle.

The response function gives the distribution of primary energy of nuclei of mass $A$
that produce leptons of a given energy $E_\ell$ as
\begin{equation}
R_\ell(A,E,E_\ell,\theta)\,=\,\phi_N(A,E)\times \frac{{\rm d}N_l(>E_l,A,E,\theta)}{{\rm d}E_l}.
\label{response}
\end{equation}
Then the flux of leptons is
\begin{equation}
\phi_\ell(E_\ell,\theta)\,=\,\Sigma_A\,\int{\rm d}E\,R_\ell(A,E,E_\ell,\theta).
\label{flux}
\end{equation}

To estimate the passing rate of neutrinos we evaluate
\begin{equation}
P_\nu(E_\nu,\theta)\,=\,\frac{\Sigma_A\,\int\,{\rm d}E\,R_\nu
P(N_{\mu} = 0)}{\Sigma_A\,\int\,{\rm d}E\,R_\nu},
\label{passing-approx}
\end{equation}
where $P(N_{\mu} = 0 | A,E,E_{\mu,{\rm min}},\theta)$ is the probability that
no muons from a shower initiated by a cosmic ray of the given mass, energy, and
zenith angle penetrate to the depth of the detector without dropping below the
detection threshold $E_{\mu,{\rm min}}$. This can be approximated as the Poisson
probability
\begin{equation}
	P(N_{\mu} = 0 | E,E_{\mu,{\rm min}},\theta) = e^{-N_\mu(A,E,\tilde{E}_{\mu,{\rm min}}(\theta),\theta)},
\end{equation}
where $\tilde{E}_{\mu,{\rm min}}(\theta)$ is the surface energy required to
reach the detector with $E_{\mu,{\rm min}}$ 50\% of the time and $N_\mu$ is the
cumulative muon yield evaluated at that energy. A Python implementation of this
calculation is included with the online supplemental materials for this article.

The central idea of the estimate is to weight the probability of zero muons
according to the weights that give rise to the flux of neutrinos of a given
$E_{\nu}$. 
It should be noted that the same idea can be applied to estimate the
atmospheric neutrino veto efficiency of a surface detector like
IceTop~\cite{IceTop} by replacing $P(N_{\mu} = 0 | A,E,E_{\mu,{\rm min}},\theta)$
with $1-\epsilon(A,E,\theta)$ where $\epsilon$ is the surface detector's trigger
efficiency for showers initiated by a cosmic rays of the given mass, energy, and
zenith angle.

\section{Results}
\label{sec:results}

\begin{figure}
\includegraphics[scale=1]{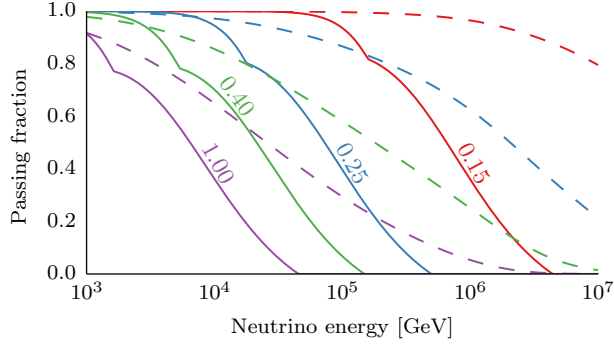}
\caption{(Color online) Solid lines show the passing rate as a function of neutrino energy from the
analytic calculation of Ref.~\protect\cite{Schoenert:2009} for various $\cos\theta$.  Dashed lines show the
calculation of Eq.~\ref{passing-approx}. The passing rate in both calculations
increases rapidly with the depth of the detector.}
\label{comparePassing}
\end{figure}

In Fig.~\ref{comparePassing} we compare the veto passing fraction (fraction of
neutrinos that arrive at a depth of 1950 m in ice with no muons above 1 TeV) in
two cases: once considering only muons produced in the same decay as the
$\nu_{\mu}$ \cite{Schoenert:2009}, and once considering uncorrelated muons from
other branches of the shower (Eq.~\ref{passing-approx}). As might be expected,
the uncorrelated veto by itself is not as efficient as the same-decay veto, but it
applies to $\nu_e$ as well as $\nu_{\mu}$.

\begin{figure}[h]
\includegraphics[scale=1]{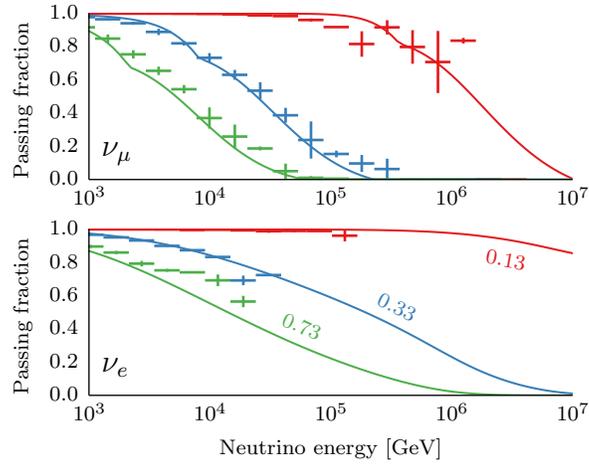}
\caption{(Color online) Comparison of approximations (solid lines) with Monte Carlo (crosses)
for conventional neutrinos at 3 values of $\cos\theta$.  
Top panel: $\nu_\mu$ with solid lines showing the passing rate from the
analytic calculation of Ref.~\protect\cite{Schoenert:2009}.  Bottom panel: $\nu_e$ with solid lines
showing the approximate calculation of Eq.~\ref{passing-approx}.}
\label{compareMC}
\end{figure}

Fig.~\ref{compareMC} shows a comparison of the analytic calculations of the
passing rate with a full Monte Carlo calculation. We simulated showers with
\textsc{corsika}~\cite{CORSIKA} and \textsc{sibyll}~2.1~\cite{SIBYLL_2.1} hadronic interactions,
weighting the showers to the H3a spectrum of Ref.~\cite{Gaisser:2012}. We then
used \textsc{proposal}~\cite{PROPOSAL} to propagate the muons in each shower through ice
to a vertical depth of 1950 m, and tabulated the fraction of neutrinos where no
muons reached depth with more than 1 TeV as a function of neutrino energy,
flavor, and zenith angle. Since $\nu_{\mu}$ may be vetoed either by a muon from
the same vertex or from the rest of the shower, we approximate the passing rate
as
\begin{equation}
P_{\rm total}\,\approx\,P_{\rm correlated}\times P_{\rm uncorrelated},
\label{combined}
\end{equation}
where the first factor is the passing rate from Ref.~\cite{Schoenert:2009} and
the second is from Eq.~\ref{passing-approx}. While this approximation accounts for the
correlated muon more than once, it nonetheless describes the full Monte Carlo calculation quite well.
For $\nu_e$ there is no partner muon, and the passing rate is described well by
Eq.~\ref{passing-approx} alone.

\begin{figure}[h]
\includegraphics[scale=1]{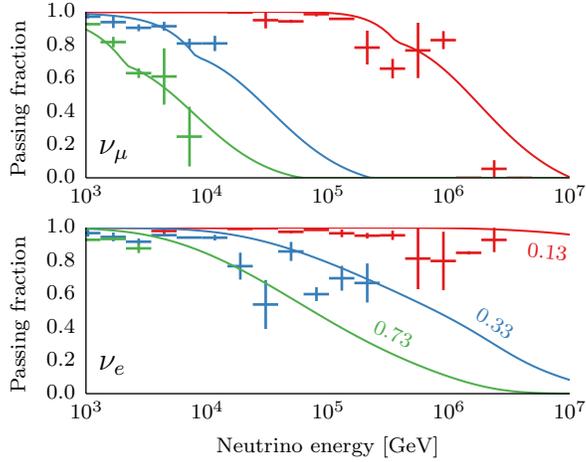}
\caption{(Color online) Comparison of approximations (solid lines) with Monte Carlo (crosses)
for neutrinos from charmed meson decay at 3 values of $\cos\theta$.  
Top panel: $\nu_\mu$ with solid lines showing the passing rate from the
analytic calculation of Ref.~\protect\cite{Schoenert:2009}. While this
calculation only applies strictly to 2-body decays of pions and kaons, it
provides an adequate description of the muon/neutrino correlation in 3-body
decays of D mesons as well.
Bottom panel: $\nu_e$ with solid lines showing the approximate calculation of Eq.~\ref{passing-approx}.}
\label{compareMCCharm}
\end{figure}

Fig.~\ref{compareMCCharm} shows a similar comparison, but only considers neutrinos
from the decays of charmed mesons simulated with \textsc{dpmjet}~2.55~\cite{DPMJETCharm}.
Here, the passing rate for $\nu_{\mu}$ is nearly the same as in the conventional
case, while for $\nu_e$ it is slightly higher. This happens because neutrinos
from charm decay tend to carry a larger fraction of the shower energy than
conventional neutrinos, and so come from a population of showers with less energy
and fewer muons on average.

\begin{figure}[hbt]
	\centering
	\includegraphics[scale=1]{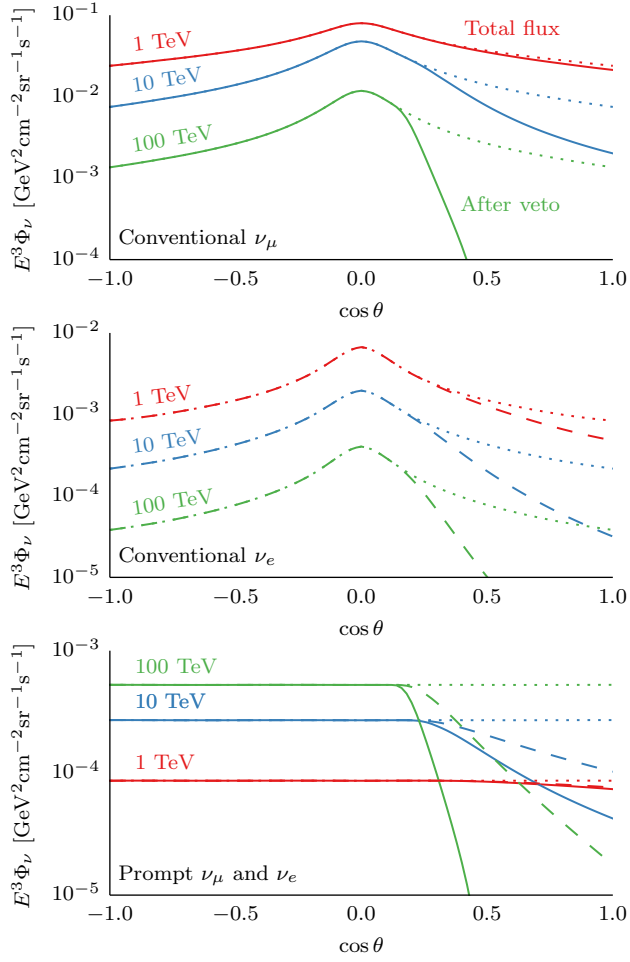}
	\caption{(Color online) Effective atmospheric neutrino flux obtained by applying the
	passing-rate calculation presented here to the conventional flux calculation
	of Ref.~\cite{Honda:2006} and the prompt flux calculation of
	Ref.~\cite{Enberg:2008}. The dotted lines in each panel show the total
	neutrino flux as a function of zenith angle for different energies, while
	the solid (dashed) lines show the portion of the $\nu_{\mu}$ ($\nu_{e}$) flux that can reach IceCube with no
	accompanying muons above 1 TeV. Above 100 TeV the up-going neutrino flux is
	suppressed by absorption in the Earth; this effect is not shown.}
	\label{fig:veto_zenith}
\end{figure}

Finally, Fig.~\ref{fig:veto_zenith} shows the effective neutrino fluxes that
can be observed in IceCube if all neutrino events with accompanying muons above
1 TeV are removed from the sample. The most notable feature is that the up-down
symmetry of the atmospheric neutrino flux is distorted for zenith angles
smaller than 75$^{\circ}$. The effect is especially stark for the prompt
neutrino flux, which would otherwise be completely isotropic, mimicking a diffuse
flux of extragalactic neutrinos.

Self-veto provides a powerful tool for disentangling astrophysical neutrinos
from an otherwise irreducible atmospheric neutrino background, and vice versa.
The generalized calculation presented here can be used to estimate passing
rates for conventional and prompt neutrinos of all flavors.

\begin{acknowledgments}

The authors are supported by grants from the National Science Foundation. The
Monte Carlo simulations were performed using the compute resources and
assistance of the UW-Madison Center For High Throughput Computing (CHTC) in the
Department of Computer Sciences. The CHTC is supported by UW-Madison and the
Wisconsin Alumni Research Foundation, and is an active member of the Open
Science Grid, which is supported by the National Science Foundation and the
U.S. Department of Energy's Office of Science.

\end{acknowledgments}

\begin{table}
	\centering
	\caption{Parameters of the modified Elbert formula for different lepton flavors and production processes.}
	
	\begin{tabular}{l D{.}{.}{6} D{.}{.}{3} D{.}{.}{3} D{.}{.}{3} c }
	Parameterization           & {\rm K}   & p_1       & p_2       & p_3   & Equation    \\
	\hline                                                                  
	Elbert $\mu$               & 14.5      & 0.757     & 5.25      & 1     & \eqref{Elbert}   \\

	Conventional $\mu$         & 49.5      & 0.626     & 4.94      & 0.580 & \eqref{Elbert}   \\
	Conventional $\nu_{\mu}$   & 79.9      & 0.463     & 4.37      & 0.316 & \eqref{Elbert}   \\
	Conventional $\nu_{e}$     & 0.805     & 0.619     & 9.78      & 0.651 & \eqref{Elbert}   \\

	Charm $\nu_{\mu}$ and $\nu_{e}$       & 0.000780  & 0.604     & 7.34      & 0.767  & \eqref{ElbertCharm}   \\
	\end{tabular}
	\label{tab:elbert_params}
\end{table}

\end{document}